\documentstyle[12pt]{article}

\newcommand{\be}{\begin{equation}}
\newcommand{\ee}{\end{equation}}
\newcommand{\ba}{\begin{array}}
\newcommand{\ea}{\end{array}}
\newcommand{\bea}{\begin{eqnarray}}
\newcommand{\eea}{\end{eqnarray}}

\newcommand{\der}{\partial}

\newcommand{\ajo}[2]{\frac{\der #1}{\der #2}}

\newcommand{\half}{{1\over2}}

\newcommand{\gcal}{{\cal G}}
\newcommand{\fcal}{{\cal F}}

\newcommand{\suma}{\sum_{i=1}^{N}}
\begin{document}

\thispagestyle{empty}
\mbox{}\\[1cm]
\begin{center}
{\bf DIRAC QUANTIZATION OF TWO-DIMENSIONAL DILATON GRAVITY
MINIMALLY COUPLED TO N MASSLESS SCALAR FIELDS}
\end{center}
\vspace{0.5 cm}
\begin{center}
{\sl by\\
}
\vspace*{0.50cm}
{\bf
Domingo Louis-Martinez}

\vspace*{0.50cm}
{\sl
Department of Physics and Astronomy, University of British Columbia\\ 
Vancouver, BC, Canada, V6T 1Z1

{[e-mail: martinez@physics.ubc.ca]}\\[5pt]
}

\end{center}
\bigskip\noindent
{\large
ABSTRACT
}
\par
\noindent
It is shown that the Callan-Giddings-Harvey-Strominger theory on 
$S^{1}\times R$
can be consistently quantized (using Dirac's approach)
without imposing any constraints on the sign of the gravitational
coupling constant or the sign (or value) of the cosmological constant.
The quantum constraints in terms of the original geometrical variables
are also derived. 

\newpage

\pagenumbering{arabic}

Two-dimensional theories of gravity have been studied as simplified
models that might provide some insight into the problems that appear
in the quantization of the (3+1)-dimensional theory.

In a recent paper \cite{1} E. Benedict, R. Jackiw and H.-J. Lee
showed that 
the CGHS model \cite{2} on the cylinder ($S^{1}\times R$) can be quantized 
using Dirac's approach in the functional
Schrodinger representation. The open case ($R\times R$) was considered  
by K. Kuchar, J.D. Romano and M. Varadarajan in \cite{6}. 

In \cite{1}, a constraint on the sign of $G / \lambda$ 
(where $G$ is the 
gravitational coupling constant
and $\lambda$ the cosmological constant)
was imposed in order  
to assure the invertibility of the transformation 
that brings the Dirac constraints to the simple form
of the pure string-inspired gravity theory. 
In terms of these new variables, the exact physical quantum states 
of the model were obtained \cite{1}.

In the conclusions to \cite{1} it was pointed out that the constraint on 
the sign of $G / \lambda$ arose from considering the particular form
of the canonical transformation that was used and, therefore,
it was not expected to be generic.

In this report we show how this problem can be solved by introducing
a different canonical transformation. 
We also derive the quantum constraints in terms of the original
geometrical variables.

Let us consider the CGHS action functional \cite{1}:

\be
S = - \int dt \int_{0}^{2\pi}dx \sqrt{-g} \left(
{1 \over 2G} (\phi R + 2\lambda) + \half \sum_{i=1}^{N}
g^{\alpha\beta}\bigtriangledown_{\alpha}f_{i}
\bigtriangledown_{\beta}f_{i} \right)
\label{eq1}
\ee

In (\ref{eq1}), $g_{\mu\nu}$ is the metric tensor, $R$ is the scalar
curvature, $\phi$ is the dilaton and $f_{i}$ ($i = 1,2,...,N$) are 
minimally coupled (to the metric tensor) massless scalar matter fields.

As in \cite{1}, we assume space-time to be a cylinder ($S^{1}\times R$).

The metric can be parametrized as

\be
ds^{2} = e^{2\rho} \left( -\sigma^{2} dt^{2} + (dx + Mdt)^{2}  \right)
\label{eq2}
\ee

In this parametrization, the canonical Hamiltonian takes the form:

\be
H_c = \int_{0}^{2\pi}dx \left( M{\cal F} + \sigma{\cal G}   \right)
\label{eq3}
\ee

\noindent where,

\be
{\cal F} \equiv \rho' \Pi_{\rho} + \phi' \Pi_{\phi} 
- \Pi'_{\rho} + \sum_{i=1}^{N} f'_{i} \Pi_{i} \approx 0
\label{eq4a}
\ee

\be
{\cal G} \equiv -{\phi'' \over G} + {\phi' \over G}\rho' +
G \Pi_{\phi} \Pi_{\rho} + {\lambda \over G} e^{2\rho} +
\half \sum_{i=1}^{N} \left(\Pi_{i}^{2} + (f'_{i})^{2} \right)
\approx 0
\label{eq4b}
\ee

\noindent are the secondary first-class constraints \cite{4} of the 
system. $\Pi_{\phi}$, $\Pi_{\rho}$ and $\Pi_{i}$
are the momentum densities conjugate
to $\phi$, $\rho$ and $f_{i}$. In (\ref{eq4a},\ref{eq4b}) 
$\approx$ denotes weak equality in Dirac's sense \cite{4}. 
The momentum densities conjugate
to $M$ and $\sigma$ are the primary constraints, which are also first-class.
No other constraints appear following Dirac's algorithm.

The constraints (\ref{eq4a},\ref{eq4b}) satisfy the following Poisson-bracket
(PB) relations:

\be
\{\fcal(x), \fcal(y)\} =
\{\gcal(x), \gcal(y)\} =
\left( \fcal(x) + \fcal(y) \right) \ajo{}{x} \delta (x - y)
\label{eq5a}
\ee

\be
\{\fcal(x), \gcal(y)\} =
\{\gcal(x), \fcal(y)\} =
\left( \gcal(x) + \gcal(y) \right) \ajo{}{x} \delta (x - y)
\label{eq5b}
\ee

It is convenient to consider the following equivalent set
of first-class constraints:

\be
C_{-} \equiv \half (\fcal - \gcal)
\label{eq6a}
\ee

\be
C_{+} \equiv \half (\fcal + \gcal)
\label{eq6b}
\ee

We introduce a canonical transformation defined as follows:

\be
(X^{-})' = -{1 \over \alpha} e^{\Pi_{A}} 
\label{eq7a}
\ee

\be
P_{-} = \alpha \left( - (A e^{-\Pi_{A}})' + 
{\lambda \over 2G} e^{-\Pi_{B}} \right)
\label{eq7b}
\ee

\be
(X^{+})' = {1 \over \beta} e^{-\Pi_{B}} 
\label{eq7c}
\ee

\be
P_{+} = \beta \left( - (B e^{\Pi_{B}})' + 
{\lambda \over 2G} e^{\Pi_{A}} \right)
\label{eq7d}
\ee

\noindent where,

\be
A = \half ({\phi' \over G} - \Pi_{\rho})
\label{eq8a}
\ee

\be
\Pi'_{A} = \rho' - G \Pi_{\phi}
\label{eq8b}
\ee

\be
B = \half ({\phi' \over G} + \Pi_{\rho})
\label{eq8c}
\ee

\be
\Pi'_{B} = - \rho' - G \Pi_{\phi}
\label{eq8d}
\ee

The positive real numbers $\alpha$ and $\beta$ are defined as:

\be 
\alpha= {1 \over 2\pi} \int_{0}^{2\pi}dz e^{\Pi_{A}} 
\label{eq100}
\ee

\be
\beta= {1 \over 2\pi} \int_{0}^{2\pi}dz e^{-\Pi_{B}}  
\label{eq101}
\ee

In terms of these new variables, the constraints 
(\ref{eq6a}, \ref{eq6b}) take the simple form:

\be
C_{-} \equiv (X^{-})' P_{-} - {1 \over 4} \suma
\left( \Pi_{i} - f'_{i} \right)^{2} \approx 0
\label{eq9a}
\ee

\be
C_{+} \equiv (X^{+})' P_{+} + {1 \over 4} \suma
\left( \Pi_{i} + f'_{i} \right)^{2} \approx 0
\label{eq9b}
\ee

This is exactly the form obtained in \cite{1,6,3,5}, however, our 
canonical transformation (\ref{eq7a} - \ref{eq8d}) 
is different from the ones presented in those papers.

Notice that from (\ref{eq7a}, \ref{eq7c}) it follows that:

\be
(X^{-})' < 0
\label{eq10a}
\ee

\be
(X^{+})' > 0
\label{eq10b}
\ee

This result follows solely from the canonical redefinition 
(\ref{eq7a} - \ref{eq8d}) without
imposing any constraints on the sign of the gravitational coupling
constant $G$ or on the sign of the cosmological constant $\lambda$.
Notice also that (\ref{eq10a}, \ref{eq10b}) are valid in general 
for any field configuration, regardless of whether or not it 
satisfies the constraints. Finally,
notice that no restriction has been imposed on the value of the 
cosmological constant ($\lambda$ can take any value, including zero).

From (\ref{eq7a}, \ref{eq7c}, \ref{eq100}, \ref{eq101}) 
it also follows that

\be
X^{-}(2\pi) - X^{-}(0) = -2\pi
\label{eq11a}
\ee

\be
X^{+}(2\pi) - X^{+}(0) = 2\pi
\label{eq11b}
\ee

This (\ref{eq9a} - \ref{eq11b}) allows us to use the results of \cite{3}.

So far we have considered only the classical theory. Let us now turn
our attention to the quantization of the model (\ref{eq1}).

At the quantum level, the constraints (\ref{eq9a}, \ref{eq9b}) 
need to be modified \cite{1,3}
in order to satisfy Dirac's consistency conditions \cite{4}. The quantum
constraints $C_-$ and $C_+$ in terms of Kuchar's variables can be written
as \cite{1}:

\be
C_{-} = (X^{-})' P_{-} - 
{N \hbar \over 48\pi} \left(\ln(-(X^{-})') \right)'' 
- {1 \over 4} \suma
\left( \Pi_{i} - f'_{i} \right)^{2} + 
{N \hbar \over 48\pi}
\label{eq14a}
\ee

\be
C_{+} = (X^{+})' P_{+} + 
{N \hbar \over 48\pi} \left(\ln((X^{+})') \right)'' 
+ {1 \over 4} \suma
\left( \Pi_{i} + f'_{i} \right)^{2} -
{N \hbar \over 48\pi}
\label{eq14b}
\ee

From (\ref{eq14a}, \ref{eq14b}), using the canonical 
transformation (\ref{eq7a} - \ref{eq8d}) and the definitions 
(\ref{eq6a}, \ref{eq6b}), we can 
obtain the quantum constraints ${\cal F}$ and 
${\cal G}$ in terms of the original geometrical variables 
(\ref{eq1}, \ref{eq2}):

\be
{\cal F} \equiv \rho' \Pi_{\rho} + \phi' \Pi_{\phi} 
- \Pi'_{\rho} + 
{GN\hbar \over 24\pi}\Pi'_{\phi}
+ \sum_{i=1}^{N} f'_{i} \Pi_{i} 
\label{eq15a}
\ee

\be
{\cal G} \equiv -{\phi'' \over G} + {\phi' \over G}\rho' +
G \Pi_{\phi} \Pi_{\rho} + {\lambda \over G} e^{2\rho} +
{N\hbar \over 24\pi}\rho''
+ \half \sum_{i=1}^{N} \left(\Pi_{i}^{2} + (f'_{i})^{2} \right)
- {N\hbar \over 24\pi}
\label{eq15b}
\ee

The terms    
${GN\hbar \over 24\pi}\Pi'_{\phi}$ and  
${N\hbar \over 24\pi}\rho''$, 
added to ${\cal F}$ and ${\cal G}$ respectively, 
cancel the Schwinger terms (in the commutators between ${\cal F}$ and 
${\cal G}$) arising from the presence of N matter scalar fields 
in the model. 
The term 
${N\hbar \over 24\pi}$
substracted from the Hamiltonian constraint 
${\cal G}$ originates from the fact that space is assumed to be a circle
(Casimir effect).

The quantum constraints (\ref{eq15a}, \ref{eq15b}) satisfy 
the commutation relations:

\be
[\fcal(x), \fcal(y)] =
[\gcal(x), \gcal(y)] =
i\hbar \left( \fcal(x) + \fcal(y) \right) \ajo{}{x} \delta (x - y)
\label{eq16a}
\ee

\be
[\fcal(x), \gcal(y)] =
[\gcal(x), \fcal(y)] =
i\hbar \left( \gcal(x) + \gcal(y) \right) \ajo{}{x} \delta (x - y)
\label{eq16b}
\ee

Therefore, Dirac's quantization \cite{4} can be carried out in a 
consistent way.

The quantum canonical 
transformation \cite{1,3}:

\be
{\bar X}^{\pm} = X^{\pm}
\label{eq12a}
\ee

\be
({\bar X}^{\pm})'{\bar P}_{\pm} = C_{\pm}
\label{eq12b}
\ee

\be
Q^{(i)\pm}_{n} = {1 \over \sqrt{2\pi}} \int_{0}^{2\pi}dz
\cos(n X^{\pm}) (\Pi_{i} \pm f'_{i})
\label{eq12c}
\ee

\be
\Pi^{(i)\pm}_{n} = {1 \over \sqrt{2\pi}n} \int_{0}^{2\pi}dz
\sin(n X^{\pm}) (\Pi_{i} \pm f'_{i})
\label{eq12d}
\ee

\be
q_{i} = {1 \over \sqrt{2\pi}} \int_{0}^{2\pi}dz \Pi_{i}
\label{eq12e}
\ee

\be
p_{i} = {1 \over 2\sqrt{2\pi}} \int_{0}^{2\pi}dz 
\left(  X^{-}(\Pi_{i} - f'_{i}) + X^{+}(\Pi_{i} + f'_{i}) \right)
\label{eq12f}
\ee

\noindent brings the constraints 
(\ref{eq14a}, \ref{eq14b}) to the 
simple form of the pure ($N =0$) string-inspired gravity
theory.

In (\ref{eq12c}, \ref{eq12d}) $n = 1,2,...$ are positive integers.

The invertibility of the transformation (\ref{eq12a} - \ref{eq12f})
is guaranteed
by the conditions (\ref{eq10a} - \ref{eq11b}).
Notice that we assume $f(0)= f(2\pi)$ (periodicity condition).

The operators (\ref{eq12c} - \ref{eq12f}) commute with the quantum 
constraints (\ref{eq14a}, \ref{eq14b}). 
Therefore, (\ref{eq12c} - \ref{eq12f}) are quantum observables in Dirac's 
sense \cite{4}.

In this representation, the physical quantum states are given by the
wave functions \cite{1,3}:

\be
\chi(Q^{(i) \pm}_{n},q_{j})
\label{eq17}
\ee

{\large\bf Conclusions}

A canonical transformation which relates the original geometrical
variables for the CGHS model on $S^{1}\times R$ with variables
similar to the ones used by Kuchar \cite{3} and more recently by
E. Benedict, R. Jackiw and H.-J. Lee \cite{1} was presented.
The new variables $X^{\pm}$ automatically satisfy the 
conditions $(X^{+})' > 0, (X^{-})' < 0$
and $X^{\pm}(2\pi) - X^{\pm}(0) = \pm 2\pi$. These conditions guarantee that 
the (classical and quantum)
canonical redefinitions that bring the constraints to the simple form
of the pure dilaton gravity theory are invertible. The quantization can 
then proceed as in \cite{1,3}, without imposing any 
constraints on the sign of the gravitational coupling constant $G$
or on the sign (or value) of the cosmological constant $\lambda$.
The (modified) quantum constraints in terms of the geometrical 
variables were obtained.

\vspace{1.0cm}

\par\noindent
{\large\bf Acknowledgements}
\par
This work was supported by the National Sciences and 
Engineering Research Council of Canada.


\newpage

\begin{thebibliography}{99}

\bibitem{1}
E. Benedict, R. Jackiw and H.J. Lee, 
hep-th/9607062 (1996) (to appear in Phys. Rev. D).

\bibitem{2}C.G. Callan, S.B. Giddings, J.A. Harvey
and A. Strominger, Phys. Rev. D{\bf 45}, R1005 (1992).

\bibitem{6}K. Kuchar, J.D. Romano and M. Varadarajan,
gr-qc/9608011 (1996).

\bibitem{3}K. Kuchar, Phys. Rev. D{\bf 39}, 1579 (1989);
Phys. Rev. D{\bf 39}, 2263 (1989).

\bibitem{4}P.A.M. Dirac, {\it Lectures in Quantum Mechanics}
(Belfer Graduate School of Science, Yeshiva University,
New York, 1964).

\bibitem{5}D. Cangemi, R. Jackiw and B. Zwiebach,
Ann. Phys. (N.Y.) {\bf 245} ,408 (1996).



\end{thebibliography}
\end{document}